# Topologically protected long-range coherent energy transfer

Yujing Wang[1], Jun Ren[2], Weixuan Zhang[1], Lu He[1], and Xiangdong Zhang[1*]

[1]Key Laboratory of advanced optoelectronic quantum architecture and measurements of Ministry of Education,

School of Physics, Beijing Institute of Technology, 100081, Beijing, China

[2]College of Physics and Hebei Key Laboratory of Photophysics Research and Application,

Hebei Normal University, 050024, Shijiazhuang, Hebei, China

*Corresponding author: zhangxd@bit.edu.cn

The realization of robust coherent energy transfer with a long range from a donor to an acceptor has many important applications in the field of quantum optics. However, it is hard to be realized using conventional schemes. Here, we demonstrate theoretically that the robust energy transfer can be achieved using a photonic crystal platform, which includes the topologically protected edge state and zero-dimensional topological corner cavities. When the donor and the acceptor are put into a pair of separated topological cavities, respectively, the energy transfer between them can be fulfilled with the assistance of the topologically protected interface state. Such an energy transfer is robust against various kinds of defects, and can also occur over very long distances, which is very beneficial for biological detections, sensors, quantum information science and so on.

## 1. INTRODUCTION

Energy transfer plays an essential role in many processes in nature, such as the photosynthesis in the light-harvesting complex of green plants [1, 2]. Except for interpreting the energy transfer in nature, manipulating the energy transfer in artificial structures also has many potential applications. For example, white-light emitting nanofiber has been fabricated through the efficient energy transfer [3], sensitive and efficient biosensing has been realized by designing the novel nanostructures [4], efficient solar energy conversion can also be achieved by plasmonic nanostructures [5] and so on.

Recent years, it has been reported that the coherence may play a key role in the energy transfer [6-14], and the quantum coherence could enhance the efficiency of energy transfer [15-23]. However, under the natural circumstances, the coherence decreases quickly with the increase of the distance, thus the energy transfer efficiency decreases sharply as the distance between two molecules increases, which greatly limits its application in various aspects. How to realize long-range energy transfer with high efficiency is a challenge for us. In addition, even if the long-range energy transfer with high efficiency can be realized, how to keep them stable and free from environmental perturbations is also a problem.

New developments in the field of topological photonics [24-27] make it possible to solve the above-mentioned problems. By introducing the topology into the field of optics, numerous fascinating properties, such as the unidirectional backscattering-immune propagation of photonic edge modes [28-37], can be achieved. The combination of quantum mechanics and topology can bring more intriguing phenomena [38–48], including the topological sources of quantum light [39-41], quantum interference of topological states of light [43], topological quantum optics interface [38], topologically robust transport of entangled photons [42] and so on. Recently, a new class of higher-order topological insulators have been theoretically proposed and experimentally demonstrated in many different systems [49-68]. Moreover, the zero-dimensional (0D) topological corner state has been observed in the two-dimensional (2D) photonic crystal (PhC) slab [55, 56, 67, 68], which provides an ideal platform to design topological nanocavities [67-68].

Motivated by the above investigations, in this work we provide a scheme to realize the long-range energy transfer with a high efficiency in the PhC platform. The designed PhC platform possesses topologically protected edge state and 0D topological corner cavities. When the donor and the acceptor are put into two topological cavities, respectively, the energy transfer between them can occur through the topologically protected edge state. Such an energy transfer can always exist, even though the donor and the acceptor are separated by a long distance, and it is robust against disorders.

## 2. SCHEME AND THEORY

We consider a PhC platform, which contains five kinds of integrable modules (I-V) in different regions, as shown in Fig. 1(a). Such a platform consists of $Al_2O_3$ ($\varepsilon$=7.5) cylinders with various radius embedded in air background. All modules possess the triangular lattice with the lattice constant $a$ being 1.25μm and the unit cell contains six $Al_2O_3$ cylinders. In order to show the system more clearly, the color of cylinders in each region is different. The PhC system is assembled inside a waveguide with two pairs of high-reflectivity metallic plates placed in the $xy$ and $yz$ planes and the height of these cylinders $h$ is taken as $0.1a$, as illustrated in Fig. 1(b).

Figure 1(c) shows the schematic lattice pattern in the region I. For convenience, we define that the radius of the larger (smaller) cylinders in the triangular lattice of all regions is marked with $r_a$ ($r_b$) and the distance from the cylinder center to the center of unit cell is marked with $R$. The $r_a$ ($r_b$) and $R$ in five modules are different. In the region I, the $r_a$ ($r_b$) is set as 0.14μm (0.16μm) and $R=0.371a$. The lattices in regions I/II and III/IV satisfy the inversion symmetry. In such a case, two topological corner cavities can appear at the crossing of four regions as marked by black pentagrams in Fig. 1(a) [66]. Figure 1(d) displays their eigen-energy spectrum, and a state at 133 THz inside the gap of the topological kink states appear.

In order to design the interface state whose frequency matches the resonance frequency of the corner cavity mode, the two high-reflectivity metal plates in the $xy$ plane [marked by $M_1$ and $M_2$ in Figs. 1(a) and 1(b)] are set to ensure the density of interface states of the waveguide modified (or

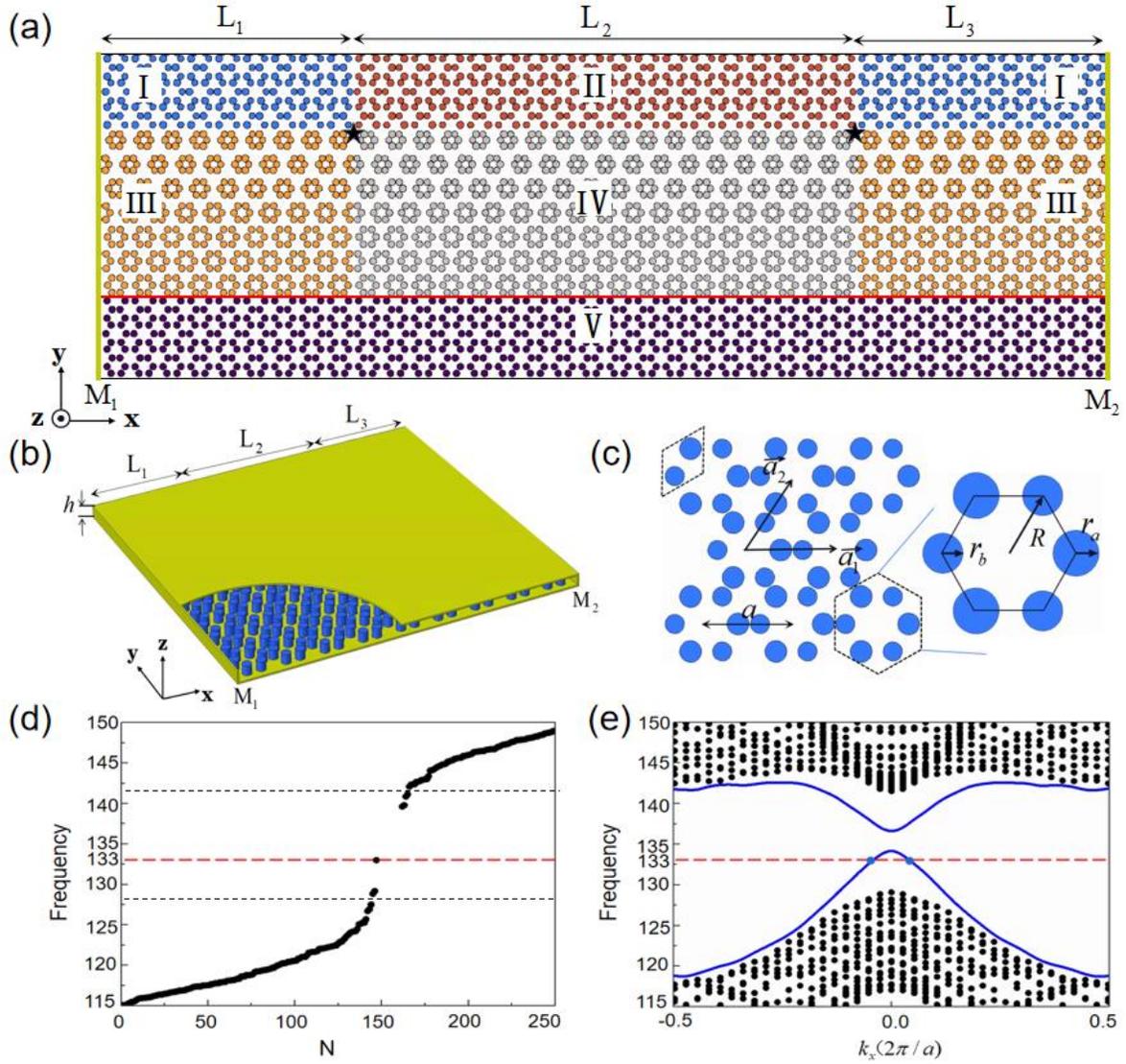

**Fig. 1.** (a) Simplified top view of the PhC system. The two black pentagrams represent two distant identical topological corner cavities (appear at the intersections of the I-IV regions), and the red line indicates a topological interface state. (b)Three dimensional photonic crystals with two pair of gold plates placed in the *xy* and *yz* planes. *h* is the height of the cylinders. (c)Schematic of a triangular photonic crystal composed of six $Al_2O_3$ cylinders ($\varepsilon$=7.5). Right panel: a magnified view of the unit cell. The radius of the hexagon (solid black line) is *R*, while the radius of the large (small) cylinder is marked with $r_a$ ($r_b$). (d)The eigen-energy spectrum of the topological corner state. (e) Calculated band diagram based on a supercell composed of region IV and V, where a periodic (absorbing) boundary condition is applied on the *x*(*y*) direction. Blue lines are for topological edge states.

discretized) to form Fabry-Perot (FP) waveguide modes. Besides, we need to design the regions III and IV skillfully, so the gradual change of parameters in different rows of the lattice is applied. Take the region III as an example, the gradually varied radius of the small (large) cylinders $r_a$ ($r_b$) is 0.14μm (0.16μm), 0.1425μm (0.1575μm), 0.145μm (0.155μm), 0.1475μm (0.1525μm) and 0.15μm (0.15μm) from the first row to the fifth row, and *R* is set as 0.243*a*, 0.243*a*, 0.262*a*, 0.281*a* and 0.3*a*, respectively. The corner states we designed exist in the coexistence of sublattice symmetry breaking and lattice deformation., while the design of edge states only requires lattice deformation. If we do not do the gradual change of parameters in regions III and IV, the frequency of the corner state will fall in the gap of the interface state or the bulk state, where the coupling between the corner state and the interface state can't occur [66]. In addition, the parameters in the sixth row and the seventh row are the same as the fifth row : $r_a=r_b$=0.15μm, R=0.3*a* (*a*/R>3). The $r_a=r_b$=0.15μm and R=0.36*a* (*a*/R<3) are taken in the region V. In such a case, a topologically protected interface state with the working frequency located at 133 THz, which corresponds to the resonance frequency of the corner cavity mode, can be obtained [36, 37]. This can be demonstrated by the calculated dispersion relation presented in Fig. 1(e). This gap is a result of the avoided crossing of the interface states caused by their interaction due to the broken $C_6$ symmetry at the conducting interface. Theoretical and experimental results show that the topological interface mode is still extremely robust and the backscattering can be ignored [37, 69-71 ]. The distance between two topological corner cavities is $L_2$, and the distance between the left (right) cavity and the left (right) metallic plates is $L_1$ ($L_3$). Here, we consider the transverse magnetic mode with the out-of-plane

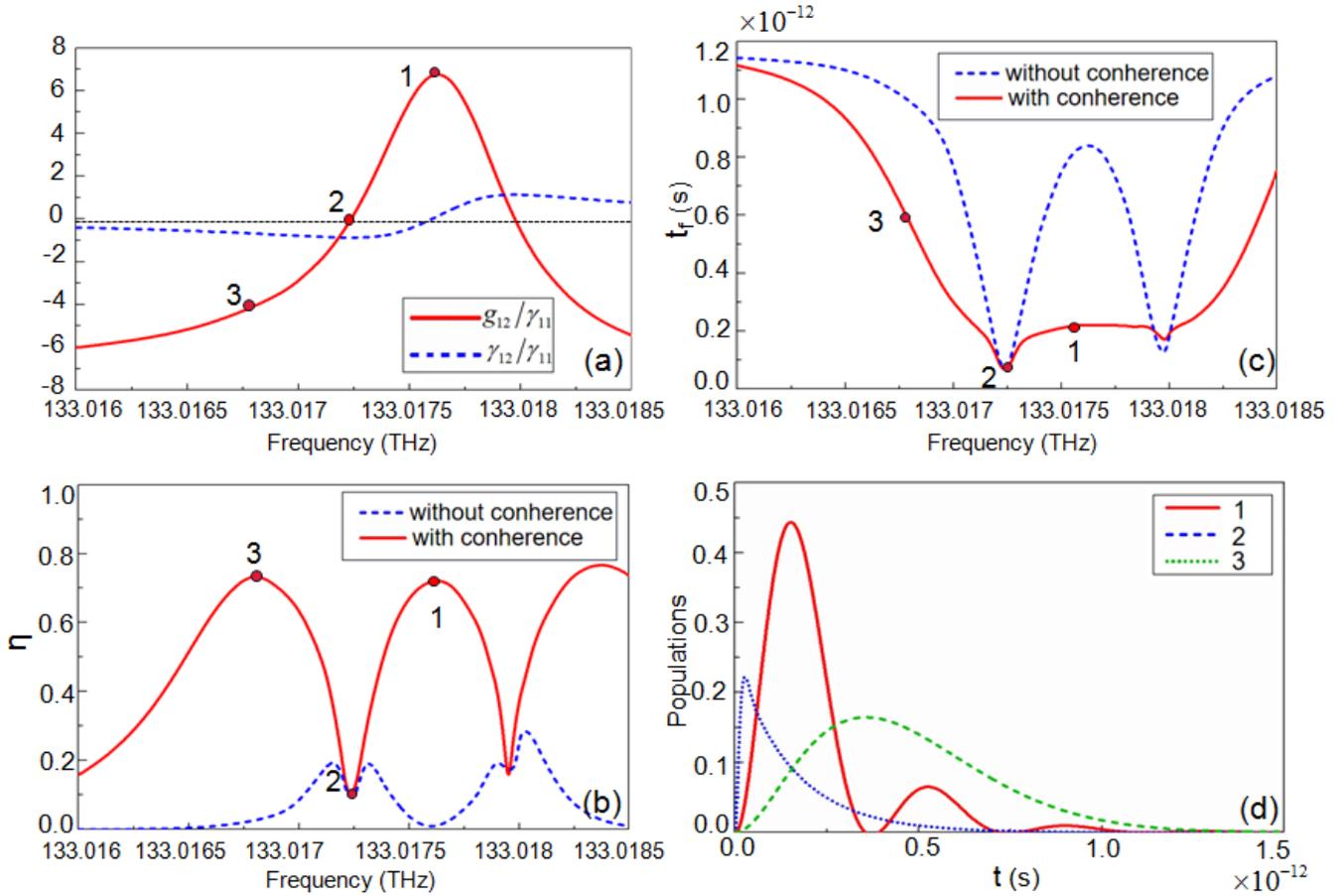

Fig. 2. (a) Coherent (red solid curve) and incoherent (blue dashed curve) couplings between two molecules with the parameters $L_2=20a$ and $L_1=L_3=19a$. Energy transfer efficiency $\eta$. (b) and energy transfer time $t_f$ (c) as a function of the frequency with (red solid line) and without (blue dashed lines) coherent coupling. (d) Populations of acceptor under three typical frequencies marked in (a), (b) and (c).

electric field and the in-plane magnetic field.

We put a pair of two-level molecules into the topological corner cavities, respectively. The dynamic process and energy transfer properties between two molecules are studied through master equation approach. Initially, one of the molecules is excited and the other in its ground state, they are called donor (D) and acceptor (A) respectively. The corresponding Lindblad master equation of two molecules [19, 73-76], whose microscopic origin and derivation using the Born-Markov and rotating-wave approximations can be found in Refs. [74-76], is expressed as follows:

$$\partial_t \rho = \frac{i}{\hbar}[\rho, H] + \frac{(\gamma_D + 2\gamma_0)}{2}\left(2\sigma_D \rho \sigma_D^\dagger - \sigma_D^\dagger \sigma_D \rho - \rho \sigma_D^\dagger \sigma_D\right)$$
$$+ \frac{(\gamma_A + 2\kappa)}{2}\left(2\sigma_A \rho \sigma_A^\dagger - \sigma_A^\dagger \sigma_A \rho - \rho \sigma_A^\dagger \sigma_A\right)$$
$$+ \frac{\gamma_{12}}{2}\left(2\sigma_D \rho \sigma_A^\dagger - \sigma_A^\dagger \sigma_D \rho - \rho \sigma_D^\dagger \sigma_A\right)$$
$$+ \frac{\gamma_{21}}{2}\left(2\sigma_A \rho \sigma_D^\dagger - \sigma_D^\dagger \sigma_A \rho - \rho \sigma_A^\dagger \sigma_D\right)$$
(1)

where $\rho$ is the density matrix of two molecules, and $\sigma_{D(A)}^\dagger$ and $\sigma_{D(A)}$ are the creation and annihilation operators for the donor (acceptor). $\gamma_0$ is the natural dissipation in the donor and $\kappa$ is the separation rate in the acceptor, and in this work we have taken $\gamma_0 =1$ ns$^{-1}$ and $\kappa =4$ ps$^{-1}$ in our calculations [19]. Due to the symmetry of two cavities, we assume the decay rates of two molecules are the same, namely $\gamma_D = \gamma_A = \gamma_1$. The $\gamma_{12}$ represents the incoherent coupling term (interference term) between the two molecules, and we also have $\gamma_{12} = \gamma_{21}$. The $\gamma_D$ and $\gamma_{12}$ can be calculated as

$$\gamma_D = \frac{2\omega_0^2}{\varepsilon_0 c^2 \hbar} \text{Im}\left[\vec{\mu}_D^* \cdot \ddot{G}(\vec{r}_D, \vec{r}_D, \omega) \cdot \vec{\mu}_D\right], \quad (2)$$

and

$$\gamma_{12} = \frac{2\omega_0^2}{\varepsilon_0 c^2 \hbar} \text{Im}\left[\vec{\mu}_A^* \cdot \ddot{G}(\vec{r}_A, \vec{r}_D, \omega) \cdot \vec{\mu}_D\right]. \quad (3)$$

In Eqs. (2) and (3), $\ddot{G}(\vec{r}_A, \vec{r}_D, \omega)$ is the classical Green tensor of the system, which is the solution of the tensor equation $[\nabla \times \nabla \times - k_0^2 \varepsilon(\vec{r}, \omega)]\ddot{G}(\vec{r}, \vec{r}'; \omega) = \bar{I}\delta(\vec{r} - \vec{r}')$, and $\vec{r}_D$ and $\vec{r}_A$ are position vectors of the donor and acceptor respectively. $\vec{\mu}_D$ and $\vec{\mu}_A$ are dipole moments of two molecules, and in this work we have taken $|\vec{\mu}_D| = |\vec{\mu}_A| = 6.4 \times 10^{-23}$ C·m. The Hamiltonian of two molecules showed in Eq. (1) can be expressed as

$$H = \sum_{i=D,A}\hbar(\omega_0 + \delta_i)\sigma_i^\dagger \sigma_i + g_{12}\left(\sigma_D^\dagger \sigma_A + \sigma_A^\dagger \sigma_D\right), \quad (4)$$

where $g_{12}$ is the coherent coupling between two molecules, which can be calculated with

$$g_{12} = \frac{\omega_0^2}{\varepsilon_0 \hbar c^2} \text{Re}\left[ \vec{\mu}_A^* \cdot \ddot{G}^s(\vec{r}_A, \vec{r}_D, \omega) \cdot \vec{\mu}_D \right], \quad (5)$$

The Green tensor $\ddot{G}^s(\vec{r}_A, \vec{r}_D, \omega)$ showed in Eq. (5) is the scattering Green tensor of the field dipole located at $\vec{r}_A$ and the source dipole located at $\vec{r}_D$, which can be obtained with

$$\vec{n}_A \cdot \ddot{G}^s(\vec{r}_A, \vec{r}_D, \omega) \cdot \vec{n}_D = -\vec{n}_A \cdot \vec{E}_s(\vec{r}_A)|_{\vec{r}_D}, \quad (6)$$

where $\vec{n}_A$ and $\vec{n}_D$ represent the direction of the field dipole and source dipole moment respectively, and $\vec{E}_s(\vec{r}_A)|_{\vec{r}_D}$ represents the scattering electric field at the location of field dipole $\vec{r}_A$ induced by the source dipole located at $\vec{r}_D$.

In this work, the orientation of the source dipole moment is taken as the positive direction of the z-axis, which will excite the TM modes of the designed structure, and the electric direction of the field dipole is also parallel to the z-axis. In this case, the above equation turns to

$$G_{zz}^s(\vec{r}_A, \vec{r}_D, \omega) = -\vec{E}^s(\vec{r}_A)|_{\vec{r}_D}, \quad (7)$$

Note that an important Markov approximation has been employed in the derivation of Eq. (1), which requires that the coherent couplings between single molecule and electromagnetic modes are much weaker than the couplings between molecules through electromagnetic field. In present work, the main conclusion is obtained when the coherent coupling between the two molecules is much larger than the coupling between the single molecule and electromagnetic field, which satisfies the Markov approximation. Thus, Eq. (1) is validity in this context. Based on Eq. (1) and the calculated parameters showed in Eqs. (2), (3) and (5), we can derive the dynamics of two molecules and investigate the coherent energy transfer between them. There are two important parameters to measure the properties of coherent energy transfer, that is, energy transfer efficiency $\eta$ and transfer time $t_f$ [19, 23, 77]

$$\eta = 2\int_0^\infty \kappa P_A(t) dt, \quad (8)$$

and

$$t_f = \int_0^\infty t P_A(t) dt / \int_0^\infty P_A(t) dt. \quad (9)$$

The energy transfer efficiency $\eta$ represents the total probability of excitation used for charge separation, and the energy transfer time means the average waiting time before charge separation happens in the acceptor [19, 23, 77]. In Eqs. (8) and (9), $P_A(t)$ is the population of the acceptor and represents the probability density of the donor in the ground state and the acceptor is in the excited state at time $t$, which is equal to the matrix element $\rho_{22}(t)$ obtained from Eq. (1) under the normal basis ({$|00\rangle$, $|01\rangle$, $|10\rangle$, $|11\rangle$}). Based on Eqs. (8) and (9), we can calculate the coherent energy transfer efficiency and time in the designed system. Note that Eqs. (8) and (9) are derived from Eq. (1), it can be seen that both of them depend not only on the parameters of the acceptor but also on the parameters of the donor.

## 3. RESULTS AND DISCUSSION

In this section, we present the calculated results for the coherent energy transfer efficiency and time in our designed PhC platform. One of the advantages of this platform is that the coherent strong coupling is easy to be realized. The red solid line in Fig. 2(a) displays the calculated results of the normalized coherent coupling coefficient ($g_{12}/\gamma_{11}$) as a function of the transition frequency. The corresponding incoherent coupling coefficient ($\gamma_{12}/\gamma_{11}$) is shown by blue dashed lines. It is seen clearly that the very strong coherent coupling ($g_{12}/\gamma_{11}$ =7.005) can be reached around the frequency $f$=133.0175 THz, at the same time, the incoherent coupling is almost equal to zero. Under the strong coherent coupling regime, if one of the molecules is excited, the excitation will be transferred back and forth between the two molecules, namely coherent energy transfer. The energy transfer efficiency $\eta$ and time $t_f$ are calculated as a function of the frequency for the case of $L_2$= 20a, the results are presented in Fig. 2(b) and 2(c), respectively. It is found that, when the coherent coupling is maximum, the energy transfer efficiency can reach 0.75 (corresponding to the mark "1" in Fig. 2(a), 2(b) and 2(c)), which is much larger than those in other systems [15-23]. At the same time, the energy transfer time is about 200 fs, namely the fast and high-efficiency coherent energy transfer is demonstrated in the present PhC platform. In fact, for energy transfer, high transfer efficiency and short transfer time have always been our goal, which is beneficial to sensors, quantum information science and so on.

We notice that there are other two peaks in the energy transfer efficiency, where both the coherent and incoherent couplings are not zero, however, at these two frequencies the energy transfer times are much larger than that at the coherent coupling peak. The reason can be found in Fig. 2(d), where the populations of the acceptor have been plotted at three typical frequencies that have been marked in Fig. 2(a)-2(c). In the case of maximal coherent coupling (marked by "1"), the population of the acceptor (red solid curve) reaches a relatively large value at the beginning. This is because the coherent coupling is proportional to the height of the quantum beat and the beat frequency, while incoherent coupling controls the duration of a single quantum beat. Therefore, the maximal coherent coupling and minimal incoherent not only guarantees larger energy transfer efficiency, but also ensures the less energy transfer time. For another efficiency peak (marked by "3"), although the integral of population over time is also large, however, it is clear that the energy transfer has also undergone a much longer time. The blue dashed curve represents the case that the incoherent coupling is maximum and coherent coupling is zero (marked by "2"). Obviously, small coherent coupling leads to a very low quantum beat, which results in a small energy transfer efficiency.

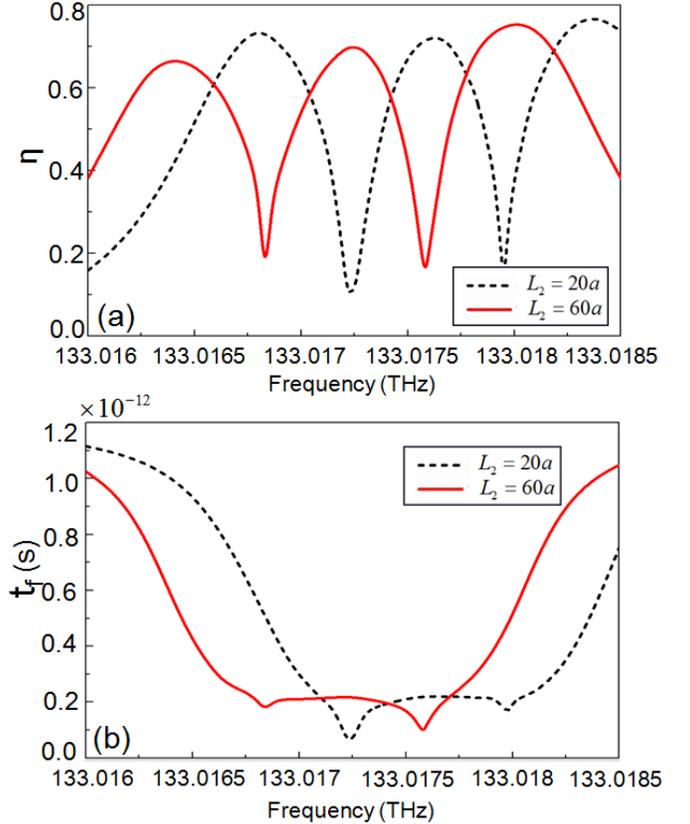

**Fig. 3.** Energy transfer efficiency (a) and transfer time (b) between two molecules separated by a longer distance $L_2$=60a (red solid lines). The black dashed lines represent the case of $L_2$=20a.

In order to explore the role of coherent coupling played in the energy transfer, we plot the results without coherence ($g_{12}=0$) by blue dashed curves in Fig. 2(b) and 2(c). It is clearly shown that, the energy transfer efficiency with coherent coupling is always larger than that without coherent coupling, and the energy transfer time is always shorter than that without coherent coupling. That is to say, coherent coupling always helps increase energy transfer efficiency while reducing energy transfer time.

The important feature of this platform is that the realized energy transfer can maintain a very long distance. The red lines in Fig. 3(a) and 3(b) represent the calculated results of the coherent energy transfer efficiency and time for the case with $L_2= 60a$. For comparison, the corresponding results for the case with $L_2= 20a$ are shown by black dashed lines. Obviously, the increase in distance only allows a slight decrease in the efficiency and a slight increase in the transfer time, except for the frequency shift presented. The frequency shift is due to the different intrinsic eigenfrequencies of different structures. This means that the present structure can be used to achieve efficient and fast energy transfer over longer distances. This is in contrast to the conventional case, where the energy transfer efficiency will decrease sharply as the increase of the distance between the donor and the acceptor, and efficient and rapid energy transfer only occurs at a small separation distance.

Another important feature of this platform is that the realized energy transfer is topologically protected, that is, it is robust against various kinds of defects. In order to exhibit such a phenomenon, we consider two types of defects. One is to introduce intentionally deformed artificial atoms at the interface, the other is to use a sharply bent interface. Here $L_2=20a$ is taken. The red solid lines in Fig. 4(a) and 4(b) display calculated results for the energy transfer efficiency and time with the first kind of defect, and the corresponding results for the second kind of defect are given in Fig.4(d) and (e). For the first defect, as shown in the black dashed rectangle of Fig. 4(c), the black cylinders represent their position has moved $0.02a$ to the left, and the white cylinders represent the radius of them has reduced to 0.145 μm. For the second defect, the original interface has been replaced by an interface with 60° and 120° turns, as shown in Fig. 4(f). For comparison, the corresponding results without defects are shown by black dashed lines. It can be seen that under the two defects, there is only a slight change in energy transfer efficiency and frequency shift, and the energy transfer time has a

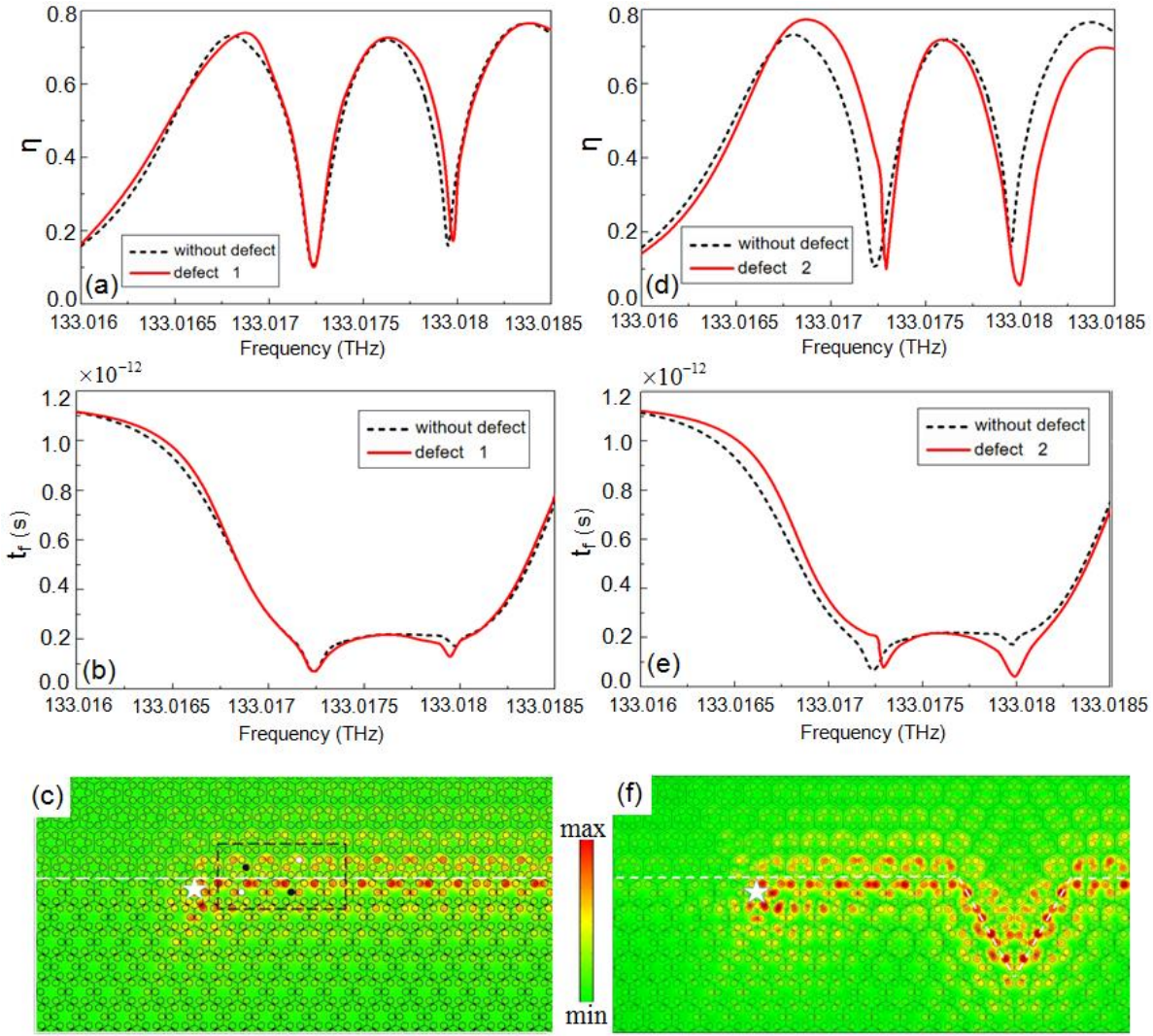

**Fig. 4.** Topologically protected energy transfer. (a), (b) and (c) correspond to the case of the first kind of defect, (d), (e) and (f) to the second kind of defect. (a) and (d) are the energy transfer efficiency (solid red lines) under the two cases, and (b) and (e) are the corresponding energy transfer time (solid red lines). The dashed black curves represent the case without defect. (c) and (f) display simulated unidirectional propagations of excited interface state along interfaces with disorders (deformed hexagons inside black dashed box) and 60° and 120° turns, respectively. The white star represents the excited chiral light source.

___

slight increase, which exhibit very good robust properties. In order to disclose the origin of these phenomena, the distribution of the electric fields at the interface with defects are plotted in Fig. 4(c) and 4(f) when the chiral light sources (marked with stars) are put on the interface. It is found that there is almost no backscattering with disorders. It is the topologically protected channel that leads to the robust energy transfer. Note that we put the source at the position of the interface not corner cavity just because we want to prove that the edge state is topologically protected. In fact，the energy transfer between the two corner cavities is topologically protected, which depends on the robustness of the interface states.

The above results are obtained without considering the effect of scattering loss due to roughness of the cylinders and absorption of the material. In fact, strong scattering loss and absorption do have influence on the above phenomena [78-79]. However, when the absorption and scattering losses are small, the effect is also small. In the wavelength range we are concerned above, the absorption of $Al_2O_3$ cylinders is quite small or even negligible, which has limited effect on the results of energy transfer. Besides, two ends of the designed photonic crystal are metal plates, which make sure that the electromagnetic waves are difficult to scatter into the air. In addition, compared with some traditional methods to realize the interaction between two quantum dots, our method can achieve a longer distance. For example, long distance 1.4μm interaction of two different InAs/GaAs quantum dots in a photonic crystal micro-cavity has been realized [72]，which is much shorter than the distance realized (tens to hundreds of microns) in our work. More importantly, long range energy transfer using this scheme is robust against various kinds of defects. Therefore, the design has more obvious advantages compared with the traditional ones, which is very beneficial for practical applications.

## 4. SUMMARY

Based on the PhC platform with the 1D topologically protected edge state and 0D corner cavities, we have demonstrated that the long-range coherent energy transfer from a donor to an acceptor can be realized. Because the donor and the acceptor have been put into topological cavities, and the energy transfer between them has been fulfilled with the assistance of the topologically protected interface state, such an energy transfer is robust against various disorders and can also occur over very long distances. These results are very important for various applications, such as biological detections, sensors, quantum information science and so on.

**Funding.** National key R & D Program of China (2017YFA0303800) and the National Natural Science Foundation of China through Grant Nos. 91850205 and 11904078. J.R was also supported by Hebei NSF (A2019205266) and Hebei Normal University (L2018B02).

**Disclosures.** The authors declare no conflicts of interest.